\documentclass[12pt]{article}
\usepackage{epsfig}
\newcommand{\be}{\begin{equation}}
\newcommand{\ee}{\end{equation}}

\newcommand{\ce}{\begin{center}}
\newcommand{\nc}{\end{center}}
\newcommand{\ed}{\end{document}}
\newcommand{\ax}{\vec{x}}

\makeatletter
\@addtoreset{equation}{section}
\makeatother

\def\sqr#1#2{{\vcenter{\vbox{\hrule height.#2pt
 \hbox{\vrule width.#2pt height#1pt \kern#1pt
 \vrule width.#2pt} \hrule height.#2pt}}}}

\def\operp{\hbox{${\kern+.25em{\bigcirc}
\kern-.85em\bot\kern+.85em\kern-.25em}$}}

\def\lsim{\;\raise0.3ex\hbox{$<$\kern-0.75em\raise-1.1ex\hbox{$\sim$}}\;}
\def\gsim{\;\raise0.3ex\hbox{$>$\kern-0.75em\raise-1.1ex\hbox{$\sim$}}\;}

\def\ce{\centerline}
\def\ve{\vfill\eject}
\def\rdots{\mathinner{\mkern1mu\raise1pt\vbox{\kern7pt\hbox{.}}\mkern2mu
 \raise4pt\hbox{.}\mkern2mu\raise7pt\hbox{.}\mkern1mu}}

\def\e e{$e^+ e^-$ }

\begin{document}

\centerline
{\bf A POSSIBLE  $SL_q(2)$  SUBSTRUCTURE OF THE STANDARD MODEL }
\vskip 0.5in

\centerline{ ROBERT J. FINKELSTEIN }

\centerline{Department of Physics and Astronomy}

\centerline{University of California, Los Angeles, CA  90095-1547, 
USA}
\centerline{ finkelstein@physics.ucla.edu }
\vskip 0.5in

\noindent{\bf ABSTRACT:}
We examine a quantum group extension of the standard model with the
symmetry  $SU(3) \times SU(2) \times U(1)\times $ global  $SLq(2)$.  The quantum fields of
this extended model lie in the state space of the  $SLq(2)$ algebra.  
The normal modes or field quanta carry the factors  $D^j _{mm^\prime}
(q|abcd)$,  which are irreducible representations of  $SLq(2)$  (which
is also the knot algebra).  We describe these field quanta as quantum
knots and set  $(j,m,m^\prime)= 1/2 (N,w, \pm r+1)$ where the  $(N,w,r)$  are
restricted to be (the number of crossings, the writhe, the rotation)
respectively, of a corresponding classical knot.

There is an empirical one-to-one correspondence between the four quantum trefoils
and the four families of elementary fermions, a correspondence that may be
expressed as  $(j,m,m^\prime )=3(t,-t_3, -t_0)$, where the four quantum
trefoils are labelled by  $(j,m,m^\prime )$  and where the four families are
labelled in the standard model by the isotopic and hypercharge indices
$(t,t_3,t_0)$.  We propose extending this correlation to all
representations by attaching  $D_{-3t-3t_0}^{3t} ( q| abcd) $ to the
field operator of every particle labelled by  $(t,t_3, t_0)$  in the
standard model.  Then the elementary fermions  $(t=1/2)$  belong to the 
$j=3/2$  representation of  $SLq(2)$.  The elements of the fundamental
representation  $j=1/2$  will be called preons and  
$D_{-3t,-3t_o}^{3t}$  may be interpreted as describing the creation
operator of a composite particle composed of elementary preons.  $D_{m m^\prime}^j$ may also
be interpreted to describe a quantum knot when expressed as   
$D_{\frac w2 \frac{\pm r+1}2} ^{N/2}$  These complementary descriptions
may be understood as describing a
composite particle of  $N$ preons bound by a knotted boson field with  
$N$  crossings.

\vskip.3cm
\noindent
{\bf UCLA/09/TEP/58}


\newpage
\noindent
\section{Introduction}

The general notion that the elementary particles are topologically
stabilized has had a long history beginning with Kelvin.${}^1$  In
recent times a classical knot related to the Skyrme soliton has been
described by Faddeev and Niemi.${}^2$
There is also an apparently unrelated preon literature beginning with
the work of Pati and Salam.${}^3$  It turns out, however, that the
knot and preon conjectures are not unrelated but may be formulated as
complementary field and particle expressions of  $SLq(2)$ symmetry.

To review the knot conjecture first note that the familiar knots of
magnetic fields are macroscopic manifestations of the electroweak
field.   It is then natural to consider knots of electroweak field that
are microscopic and quantized as well.  Since these would be observed as
solitonic in virtue of both their topological and quantum stability, it
is also natural to ask if the known elementary particles might also be
quantized knots of field.   If they are, one expects that the most
elementary particles, namely the elementary fermions, are also the most
elementary quantum knots, namely the quantum trefoils.  This possibility
is suggested by the fact that there are 4 quantum trefoils and 4 classes
of elementary fermions, and is supported by a unique one-to-one
correspondence between the topological characterization of the 4 quantum
trefoils and the quantum numbers of the 4 fermionic classes.  To define
a quantum knot we shall first record the irreducible representations of
the knot algebra $(SLq(2))$.

\section{Irreducible Representations of the Knot Algebra${}^{4,13}$}

The  $2j+1$ dimensional representation of  $SLq(2)$  may be written as
follows:

\be
D^j_{mm^\prime}(a,b,c,d) = \sum_{\scriptstyle 0\leq s\leq n_+
\atop\scriptstyle 0\leq t\leq n_-}
{\cal A}^{j}_{mm^\prime}(q,s,t)
\delta(s+t,n^\prime_+)
a^s b^{(n_{+}-s)}c^{t} d^{(n_{-}-t)}
\ee

\noindent
where
\be
n_\pm = j \pm m  
\ee
\be
n_\pm ^\prime = j \pm m'  
\ee
and the arguments  $(a,b,c,d)$  satisfy the knot algebra:${}^5$

\[
\begin{array}{llllr}
ab= qba & bd=qdb & bc = cb & ad-qbc=1 \\
ac =qca & cd=qdc & & da-q_{1}cb=1 \hspace{1.25in}\mbox{(A)}
\end{array}
\]

\noindent
where  $q_1 = q^{-1}$.

The  ${\cal A} _{mm'}^j $  are  $q$-deformations of the Wigner coefficients
that appear in irreducible representations of  $SU(2)$.

The knot algebra  (A)  and hence  $D_{mm'}^j (a,b,c,d)$  are defined
only up to the gauge transformation

\be
U_a (1): \begin{array}{l}
a^\prime =e^{i\varphi_a} a \\
d^\prime = e^{-i\varphi _a}d 
  \end{array}     \;\;\;\;\;\;\;\;\;\;\;\;\;\;
U_b (1): \begin{array}{l}
b^\prime =e^{i\varphi_b}b \\
c^\prime = e^{-i\varphi _b} c 
  \end{array}      
\ee

Eqns (2.4) leave the algebra  (A)  invariant and induce on the elements
of every representation the following  $U_a (1) \times U_b(1)$  gauge
transformation${}^6$

\be
D_{mm'}^{j}(a^{\prime}, b^{\prime}, c^{\prime}, d^{\prime})= e^{i(\varphi_a + \varphi _b)m} \;  
e^{i(\varphi _a - \varphi _b) m^{\prime}} \; D _{mm'}^{j}(a,b,c,d) 
\ee

\section{Quantization of the Knot}

Following the example of the quantization of angular momentum by
representations $D_{mm'}^j$ of $SU(2)$  where the indices  
$(j,m,m')$  refer to components of the angular momentum, we shall
quantize the kinematics of the knot with representations of the knot
algebra, also denoted by $D^j_{mm^\prime}$,
where the indices on  $D_{mm'}^j $  are now related to  
$(N,w,r)$,  the number of crossings, the writhe, and the rotation of
the corresponding classical knot,$^7$ by

\be
\begin{array}{l}
j=\frac{N}{2} \\  m= \frac{w}{2} \\  m'= \frac{\pm r+1}{2}
\end{array}
\ee

\noindent
The relations (3.1) satisfy the following restrictions:
\begin{description}
\item[(a)]  Of the set  $(N,w,r)$  only  $N$  is never negative and
therefore corresponds to  $j$  which is also never negative.
\item[(b)] Half-integer representations require the factor  $1/2$.
\item[(c)] $(2m)$  and  $(2m')$, belonging to the same representation,  
are of the same parity, while the knot constraints require  $w$  and  
$r$  to be of opposite parity.
\end{description}

Since the spectra of  $(j,m,m')$  are restricted by  $SLq(2)$,  and the
spectra of  $(N,w,r)$  are restricted by knot topology, the
states of the quantized knot are thus jointly restricted by both  
$SLq(2)$  and the knot topology.  The equations  (3.1)  then establish
a correspondence between a quantized knot described by  $D_{ \frac w2
\frac{\pm r+1}2  } ^ {N/2}$  and a classical knot described by  
$(N,w,r)$, but the correspondence is not one-to-one.

Let
\be
\Psi  _{{\frac w2} \frac{\pm  r+1}2}^{N/2} = D_{\frac w2  \frac{\pm r+1}2}^{N/2} \;
\sum _n c_n |\left. n \right>     
\ee

\noindent
where $\sum c_n \left. |n\right>$  lies in the state space defined by the knot
algebra (A).  The states, $|\left. n \right>$, forming a basis in this space are
eigenstates of the commuting elements, b and c, with eigenvalues  
$\sim q^{n(6)}$.

Here  $\Psi _{\frac w2  \frac{\pm r+1}2}^{N/2}$  is intended to describe
a generic quantum knot for which the Hamiltonian has not been
specified.  In (3.2)  $D_{\frac w2 \frac{\pm r+1}2} ^{N/2}$  is a
kinematic factor that resembles the spherical harmonic factor in  
$Y_m^\ell (\theta, \varphi) R (r)$,  an eigenstate of a spherically
symmetric Hamiltonian.  

There are, as usual, important differences
between the quantum construction and its classical image.  In particular
there are only two classical trefoils:  $(w,r)=(3,2)$  and  $(-3,2)$  
while there are four quantum trefoils labelled by
$$
(j,m,m')= \left( \frac32, \frac32, \frac32 \right) , \left( \frac 32 , - \frac32 ,
\frac 32 \right) , \left( \frac 32, \frac 32 , - \frac12 \right) , \left( \frac 32 , -
\frac 32 , - \frac 12 \right)
$$
where  $m^\prime = -\frac 12 $ corresponds to  $m^\prime = \frac{-r + 1}2 $  by  (3.1) with $r=2$.

The classical trefoils  $(w,2)$  and  $(w , -2)$  are topologically not
distinguishable.  The corresponding quantum trefoils  $(w,2)$  and  $(w,-2),$  when
realized as elementary fermions, are distinguished by
different values of the hypercharge, as we shall
see.  In the following, when  $(w,r)$  refers to a quantum trefoil,  
$r$  may have either sign, and we shall write simply 
$m^\prime =\frac{r+1}{2}$.  Then there is a one-to-one correspondence
between the quantum trefoil and the 2d-projection of the 3d-classical
knot.

There are also quantum states for which  $j < \frac 32 $.  By (3.1) these
correspond to a classical image for which  $N<3$  and do not qualify as
classical knots but may be described as twisted loops.  We shall see
that these  $j< \frac 32$  states may be realized as preons.

\section{Field Theory and Charges of Quantum Trefoils}

One may construct a field theory of the quantum trefoils  
$D_{mm'}^{3/2}$  by attaching  $D_{mm'}^{3/2}$  to a standard fermion
field operator  $\psi (x)$  as follows:

\be
\Psi _{mm'}^{3/2} = \psi (x)  \ D _{mm'}^{3/2} 
\ee

\noindent
By (2.5),  the field operator  $\Psi _{mm'}^{3/2} $ also transforms under
the gauge transformations  $U_a (1) \times U_b (1) $.
If the attachment (4.1) is made consistently for both fermionic and
bosonic fields one may construct a modified standard action that is
invariant under $U_a\times U_b$, as is shown in the appendix and in
more detail in Ref. 12.  This invariance of the field action is a
physical requirement
since the relabelling of the algebra described by  (2.4)  cannot
affect the physics.  Then in view of this invariance there will be 
by Noether's theorem one
conserved charge associated with  $U_a (1)$ and a second conserved charge
associated with  $U_b (1)$.  Then by (2.5) and (3.1) these charges may be defined by

\begin{eqnarray}
Q(w) \equiv  -k_w m &=& -k_w \ \frac{w}{2} \quad \qquad w = \pm 3 
\\
Q(r) \equiv - k_r m' &=& - k_r \ \frac{r+1}{2} \qquad
r = \pm 2  
\end{eqnarray}

\noindent
and may be referred to as the writhe and rotation charges.  
Here  $k_w$
and  $k_r$  are undetermined constants with the dimensions of an
electric charge.  In terms of  $Q(w)$  and  $Q(r)$,  the  $U_a(1) \times U_b(1)$
transformations on  $\Psi _{mm'}^{3/2}$  become

\be
\Psi _{mm'}^{3/2^\prime } = e ^{\frac{-i}{k_w} Q(w) \varphi (w) }
e^{\frac {-i}{k_r} Q(r) \varphi (r)} \ \Psi _{mm'}^{3/2} 
\ee

\noindent
where  $\varphi (w) = \varphi _a + \varphi _b $  and  $\varphi (r) =
\varphi _a - \varphi _b$  by (2.5).  We next make a direct comparison
between the  $Q(w)$  and  $Q(r)$  charges of the 
four quantum trefoils and the
charge and hypercharge of the four fermion families of the standard
theory each denoted by  
$(f_1, f_2, f_3 )$  in Table 4.1,${}^4$ since we expect that the 
simplest elementary particles are quantum trefoils 
if a knot model is
plausible.  
The knot entries in the table are determined by (3.1), (4.2),
and (4.3).


\begin{table}[h]
\begin{center}
{\bf{Table 4.1}}
\end{center}
\begin{center}
\begin{tabular}{ccccc|ccccc}
\multicolumn{5}{c}{\underline{Standard Representation}} &
\multicolumn{5}{c}{\underline{Trefoil Representation}} \\
\underline{$(f_1,f_2,f_3)$} & \underline{$t$} &
\underline{$t_3$} &
\underline{$t_0$} & \underline{$Q_e$} & \underline{$(w,r)$} &
\underline{$D^{N/2}_{\frac{w}{2}\frac{r+1}{2}}$} &
\underline{$Q_w$} & \underline{$Q_r$} & \underline{$Q_w+Q_r$} \\
$(e,\mu,\tau)_L$ & $\frac{1}{2}$ & $-\frac{1}{2}$ & $-\frac{1}{2}$
& $-e$ & (3,2) & $D^{3/2}_{\frac{3}{2}\frac{3}{2}}$ &
$-k\left(\frac{3}{2}\right)$ & $-k\left(\frac{3}{2}\right)$ &
$-3k$ \\
$(\nu_e,\nu_\mu,\nu_\tau)_L$ & $\frac{1}{2}$ & $\frac{1}{2}$ &
$-\frac{1}{2}$ & 0 & (-3,2) & $D^{3/2}_{-\frac{3}{2}
\frac{3}{2}}$ & $-k\left(-\frac{3}{2}\right)$ &
$-k\left(\frac{3}{2}\right)$ & 0 \\
$(d,s,b)_L$ & $\frac{1}{2}$ & $-\frac{1}{2}$ & $\frac{1}{6}$ &
$-\frac{1}{3}e$ & (3,-2) & $D^{3/2}_{\frac{3}{2}-\frac
{1}{2}}$ & $-k\left(\frac{3}{2}\right)$ & $-k\left(-\frac{1}{2}
\right)$ & $-k$ \\
$(u,c,t)_L$ & $\frac{1}{2}$ & $\frac{1}{2}$ & $\frac{1}{6}$ &
$\frac{2}{3}e$ & (-3,-2) & $D^{3/2}_{-\frac{3}{2}
-\frac{1}{2}}$ & $-k\left(-\frac{3}{2}\right)$ &
$-k\left(-\frac{1}{2}\right)$ & $2k$ \\
\end{tabular}
\end{center}
\end{table}

In Table (4.1)  we have assumed a single value of  $k$:

\be
k_r = k_w = k 
\ee

\noindent which is also the same for all trefoils.  If we set  $k=e/3$, we find
that the four fermion families are related to the four quantum trefoils
as follows:

\begin{eqnarray}
Q_w = e t_3 \\ 
Q_r = e t_0 \\ 
Q_w + Q_r = Q_e 
\end{eqnarray}

\noindent
in agreement with the standard model where there is the independent
relation for the electric charge

\be
Q_e = e(t_3+ t_0) 
\ee

If one aligns the trefoils and the fermion families in any order
different from that in Table 4.1, one needs more than a single value of  
$k$.  It is important that we choose  $k_r = k_w$  and that we 
choose a single value of  $k$  for the four quantum trefoils.  Note that
it is also not possible to exchange  $t_3 $  and  $t_0$  in (4.6) and
(4.7).  Therefore the correspondence between the four fermion families
and the four trefoils is
empirically fixed and unique.  This complete correspondence justifies
the representation of each of the four fermion families by a quantum
trefoil.

Since this correspondence is the basis of the following development,
it is worth emphasizing that there is only one assumption on which
the interpretation of the Table 4.1 is based: namely, that there is
a single value of $k$ for both $Q_w$ and $Q_r$, and for all the four
families.  (There are in fact no grounds for choosing more than one
value of $k$.)  If we therefore postulate a unique value of $k$, that
is enough to determine a unique correspondence between the four
families and the four quantum trefoils, since only for this
unique correspondence is there strict proportionality between
$(t_3,t_0,Q_e)$ and $(Q_w,Q_r,Q_w+Q_r)$ respectively.  The value of
$k$ as $\frac{e}{3}$ 
then follows from the identification of the total
charge of the trefoil, $Q_w+Q_r$, with $Q_e$, the electric charge
of the fermion.  Then $j_\mu^e = j_\mu^w+j_\mu^\nu$.

The correspondence between the quantum trefoils and the elementary
fermions may be summarized by the following relations which may
also be read directly from Table (4.1).  

\begin{eqnarray}
t   =  \frac{N}{6} &         &t  = \frac{j}{3}   \\
t_{3}  = - \frac{w}{6} &     &   t_{3} =-\frac{m}{3} \\ 
t_0  = - \frac{r+1}6     & & t_0 =  - \frac{m'}3  \\
Q_e = - \frac e6 (w+r+1) & & Q_e = -\frac e3 (m+m')
\end{eqnarray}

\noindent
Note also that
\begin{eqnarray}
Q_e = - \frac eN \left( \frac{w + r+1}2 \right) & &  
Q_e =  -e
\frac{(m+m')}{2j} 
\end{eqnarray}

\noindent
holds for all the elementary fermions.

While  $Q,t_3$  and  $t_0$  are defined in the standard model with
respect to  $SU(2)\times U(1)$, here  $Q, t_3$  and  $t_0$  are defined with
respect to the gauge transformations  $U_a (1)\times U_b (1)$  of the knot
algebra.  They are also described by  $(w,r)$  and  $(m,m')$  as shown
in Table (4.1) and Eqs. (4.11)-(4.13).  

The kinematic factors  $D_{m,m'}^j$  labelled by quantum numbers  
$(j,m,m')$  or  $(t,t_3 ,t_0)$  are multinomials lying in the knot
algebra $(A)$ and are explicitly given by (2.1).  These multinomials are
associated with the knot  $(N,w,r)$  and, like the Jones polynomial,
label the knot.

We incorporate Eqs. (4.10)-(4.12) into (4.1) as follows

\be
\Psi ^{1/2} (t_3, t_0, n)= \ \psi^{1/2} (t_3 , t_0, n) D_{-3t_3 -3
t_0}^{3/2} |\left. n \right> 
\ee

\noindent
where  $\psi^{1/2} (t_3 , t_0 , n)$  is the quantum field of the standard
model that represents the fermion with electroweak $SU(2)\times U(1)$ quantum numbers $t=1/2$ and  
$(t_3 ,t_0 )$. Here $|n>$ lies in the state space defined by the knot algebra where  $n=0,1,2$  
labels the generation, e.g.  
$(e,\mu ,\tau )$.  Then  $D_{-3t_3 -3t_0} ^{3/2} | \left. n\right> $  may be regarded as
an ``internal state function" reminiscent of a classical knot and
providing substructure to the elementary fermion fields of the standard
model.

We shall now propose that the non-trivial correspondence embodied in
Table (4.1) and expressed by (4.15) for the elementary fermions holds
more generally in the following form

\be
\Psi _{t_3 t_0}^t (n)=  \psi (t,t_3, t_0, n) D_{-3t_3-3t_0}^{3t} |n> 
\ee

\noindent
i.e., we assume that  $(t,t_3,t_0)$  are related to  $(j,m,m')$  just as
in the special case $t=\frac12$:
$$
3t = j  \eqno(4.17j)
$$
$$
3t_3 = -m  \eqno(4.17m)
$$
$$
3t_0 = -m^\prime \eqno(4.17m^\prime)
$$

\noindent
In other words we assume that there is an underlying  $SLq(2)$ symmetry
of the elementary particles that may be expressed through the internal
state functions $  D_{mm'}^j \left. | n\right> $.  For  $j\geq 1$  not all states  
$(m,m')$  of  $D_{mm'} ^j $ are filled.  The occupied states are
labelled by $ D_{-3t_3-3t_0}^{3t}$  according to  (4.17)  and are
determined by the intersection of the electroweak  $SU(2)\times U(1)$  and
the  $SU_q(2)$  symmetries.  The  $\left. | n\right>$  are intended to represent the
possible states of excitation of the quantum knot.  The
analogue of Table 4.1 for the elementary fermions is Table 4.2 for the
elementary bosons of the Weinberg-Salam model.

\ve

\begin{table}[h]
\begin{center}
\bf{Table 4.2}
\end{center}
\begin{center}
\begin{tabular}{ccccc}
    &\underline{$t$} & \underline{$t_3$} & \underline{$t_0$} & 
                \underline{$D_{-3t_3 -3k_0}^{3t}$}\\
$W^+$  & $1$   &  $1$ & $0$  & $D_{-30}^3$ \\
$W^-$  & $1$   & $-1$  & $0$  & $D_{30}^3$ \\
$W^3$  & $1$   & $ 0$ & $0$ & $D_{00}^3$ \\
$W^0$  & $0$   & $ 0$  & $ 0$ &  $D_{00}^0$
\end{tabular}
\end{center}
\end{table}

\noindent
We adopt the following rule:

If a particle is labelled in the standard model by electroweak quantum
numbers  $(t,t_3,t_0)$  then attach to the quantum field operator of
that particle the factor  $D_{-3t_3 -3t_0}^{3t} (a,b,c,d)$.  This factor
is to be understood as an element of the $j=3t$  representation of the  
$SL_q(2)$  algebra and may be interpreted as the replacement of the point
particle of the standard model by a solitonic structure described solely
by this factor.  The extension of (4.15) to (4.16) expresses the
conservation of $t_3$ and $t_0$ everywhere in the modified standard
model as a joint consequence of the $U_a\times U_b$ 
and $SU(2)\times U(1)$ invariance.

\vskip.5cm

\section{The Electroweak Interactions${}^8$}

In the  $SLq(2)$ model the solitonic fermions interact by the emission
and absorption of solitonic bosons.  Denote the generic fermion-boson
interaction by
\be
\bar{{\cal  F}}'' \;  {\cal B}' \;  {\cal F}   
\ee

\noindent
where

\be
{\cal F} = \mbox{{\bf F}}(p,s,t_3, t_0) \left( D_{-3t_3 - 3t_0}^{3/2} \right)  |n>
\ee

\be
\bar{{\cal F}}''  = <n''| \left( \bar{D}_{-3t_3 -3t_0} ^{3/2} \right) '' \;
\bar{\mbox{{\bf F}}}''(p,s,t_3, t_0 )
\ee

\be
{\cal B}' = \mbox{\bf B}'(p,s,t_3, t_0 ) \left( D_{-3t_3-3t_0} ^3 \right)'
\ee

\noindent
and the pair $(p,s)$ refer to momentum and spin.  Then (5.1) becomes

\be
(\bar{\mbox{{\bf F}}}'' \mbox{{\bf B}}' \mbox{{\bf F}}) 
<n''| \bar{D}_{ -3t''_{3} -3t''_{0}}^{3/2} \;
D_{-3t'_3- 3t'_0}^{3} \; D_{-3t_3-3t_0}^{3/2} |n>
\ee

\noindent
The matrix elements of the standard model will then be modified by the
following form factors.

\be 
<n''| \bar{D}_{ -3t''_{3} -3t''_{0}}^{3/2} \;
D_{-3t'_3- 3t'_0}^{3} \; D_{-3t_3-3t_0}^{3/2} |n>
\ee

\noindent
Here  $n$  and  $n''$  take on the values  $0,1,2$  corresponding to the
3 generations in each family of fermions.  These form factors are
2 parameter numerical functions that are in principle observable.$^8$
To calculate them one needs the solitonic factors
$D^j_{mm^\prime}(a,b,c,d)$ shown in Tables (5.1) and (5.2).
\begin{table}[h]
\begin{center}
\bf{Table 5.1}
\end{center}
\begin{center}
\begin{tabular}{cccccc}
\underline{$(f_1,f_2,f_3)$} & \underline{$t$} &
\underline{$t_3$} &
\underline{$t_0$} & \underline{$Q$} &
\underline{$D^{3t}_{-3t_{3}-3t_{0}}$} \\
$(e,\mu,\tau)$ & $\frac{1}{2}$ & $-\frac{1}{2}$ & $-\frac{1}{2}$ &
 $-e$ & $D^{3/2}_{\frac{3}{2}\frac{3}{2}} \sim a^{3}$ \\
$(\nu_e,\nu_\mu,\nu_\tau)$ & $\frac{1}{2}$ & $\frac{1}{2}$ &
$-\frac{1}{2}$ & 0  & $D^{3/2}_{-\frac{3}{2}\frac{3}{2}} \sim c^{3}$ \\
$(d,s,b)$ & $\frac{1}{2}$ & $-\frac{1}{2}$ & $\frac{1}{6}$ &
$-\frac{1}{3}e$ &  $D^{3/2}_{\frac{3}{2}-\frac{1}{2}}
\sim a b^{2} $ \\
$(u,c,t)$ & $\frac{1}{2}$ & $\frac{1}{2}$ & $\frac{1}{6}$ &
$\frac{2}{3}e$ &  $D^{3/2}_{-\frac{3}{2} -\frac{1}{2}}
\sim c d^{2}$ \\
\end{tabular}
\end{center}
\end{table}

\begin{table}[h]
\begin{center}
\bf{Table 5.2}
\end{center}
\begin{center}
\begin{tabular}{cccccc}
    &\underline{$t$} & \underline{$t_3$} & \underline{$t_0$} & 
    \underline{$Q$} &            \underline{$D_{-3t_3 -3t_0}^{3t}$}\\
$W^+$  & $1$   &  $1$ & $0$ &$e$  & $D_{-30}^3 \sim c^3 d^3 $ \\
$W^-$  & $1$   & $-1$  & $0$ & $-e$ & $D_{30}^3 \sim a^3 b^3 $ \\
$W^3$  & $1$   & $ 0$ & $0$ & $0$ & $D_{00}^3 \sim f_{3}(b,c)$ \\
$W^0$  & $0$   & $ 0$  & $ 0$ & $0$ &  $D_{00}^0 \sim f_{0}(b,c)$
\end{tabular}
\end{center}
\end{table}

\noindent
The solitonic factors are computed according to (2.1) and are all
monomials except for the neutral  $W^0$  and  $W^3$.  The numerical
factors  ${\cal A}_{mm'}^3$  have been dropped but may be computed according to

\be
{\cal A}_{mm'}^{j}=  
\left[ \frac{\left< n_+^\prime \right>_{q_1}! \; \left<n_-^\prime\right>_{q_1}!}{\left<n_+\right>_{q_1}!\; \left<n_-\right>_{q_1}!}
\right] ^{\frac{1}{2}} \; 
\left< {\scriptstyle n_+ \atop s} \right>_{q_{1}}
\left< {\scriptstyle n_- \atop t} \right>_{q_{1}} 
\ee

\noindent
where
\be
\left< {\scriptstyle n \atop s} \right> _q
=
\frac{\left< n\right> _q !}{\left< n-s\right >_q !
\left< s\right >_q ! }  
\; \; \; \mbox{with}\; \;
\langle n\rangle _q = \frac{q^n -1}{q-1}; \;\;\;\;  q_1 =q^{-1}
\ee

Since we require that the fermion-boson interaction be expressed
by (5.1), and that the total action be invariant under both
$SU(2)\times U(1)$ and $U_a(1)\times U_b(1)$, (5.1) and (5.6)
must share this invariance.  Then since $(4.17m)$ and $(4.17m^\prime)$
hold for the elementary ${\cal{F}}$, the same equations must also
hold for ${\cal{B}}$, as noted in the Appendix.  Hence Eq. 
$(4.17m)$ and $(4.17m^\prime)$ are not simply conjectured extensions
but they are an essential requirement of the model.  We are also
imposing $(4.17j)$ but this is not required by the same argument.

Although $(N,w,r+1) = 6(t_1,-t_3,-t_0)$ is satisfied for the
elementary fermions, it does not hold in general.  For example, there
is a ditrefoil realization of this relation for the weak bosons
$W^\pm$ but not for the neutral $W^3$ and $W^0$.  There is, however,
always a $SL_q(2)$ realization by $(j,m,m^\prime) =
3(t,-t_3,-t_0)$ and
the connection to the standard model is through $(t,t_3,t_0)$.

\section{The Preon Representations.${}^4$}

The elementary fermions already discussed are found in the  $j=3/2$
representation while the electroweak bosons lie in the  $j=3 $
representation.  We shall now consider the adjoint  $(j=1)$  and
fundamental  $(j=1/2)$  representations.  These are shown in Tables
(6.1) and (6.2) again calculated with (2.1) but ignoring the numerical
coefficients.

\begin{center}
{\bf{Table 6.1}}
\end{center}
\begin{center}
\[
D^{1/2}: \qquad
\begin{array}{c|cc}
{}_m\backslash{}^{m'} & \frac{1}{2} & -\frac{1}{2} \\ \hline
\frac{1}{2} & a & b \\
-\frac{1}{2} & c & d
\end{array}
\]
\end{center}

\ve

\begin{center}
{\bf{Table 6.2}}
\end{center}
\begin{center}
\[
D^{1}: \qquad
\begin{array}{c|ccc}
{}_m\backslash{}^{m'} & 1  & 0 & -1  \\ \hline
1  & a^2 & ab &  b^2\\
0  & ac & ad +bc & bd \\
-1 & c^2 & cd  & d^2
\end{array}
\]
\end{center}

\noindent
We shall refer to the members of the  $D^{1/2}$  and  $D^1$  
representations as preons and bosonic preons respectively.

To determine  $(t_3, t_0, Q)$  for the preons and bosonic preons we
shall extend the relations  
(4.17) empirically
established for the elementary fermions, then extended to the
electroweak bosons and generally embodied in  $D_{-3t_3 -3t_0}^{3t}$.  
The results for preons and bosonic preons are shown in Tables (6.3) and
(6.4).

\vskip 0.1in

\begin{center}
{\bf{Table 6.3}} \\
{\bf Fermionic Preons} $t=1/6$
\end{center}
\[
\begin{array}{l|cccc}
 & t  & t_3  & t_0 & Q \\
 \hline
 a & \frac{1}{6} &  -\frac{1}{6}  & -\frac{1}{6} &-\frac{e}{3} \\
 b & \frac{1}{6} &  -\frac{1}{6}  & \frac{1}{6} & 0 \\
 d & \frac{1}{6} &  \frac{1}{6}  & \frac{1}{6} &  \frac{e}{3} \\
 c & \frac{1}{6} &  \frac{1}{6}  & -\frac{1}{6} &  0 
\end{array}
\]


\begin{table}[ht]
\begin{center}
{\bf{Table 6.4}} \\
{\bf Bosonic Preons} $t=1/3$
\end{center}
\begin{center}
\begin{tabular}{l|cccc||l|cccc||l|cccc}
& \underline{$t_3$} & \underline{$t_0$} & \underline{$Q/e$} &
\underline{$D^1_{mm^\prime}$} & &
\underline{$t_3$} & \underline{$t_0$} & \underline{$Q/e$} &
\underline{$D^1_{mm^\prime}$} & &
\underline{$t_3$} & \underline{$t_0$} & \underline{$Q/e$} &
\underline{$D^1_{mm^\prime}$} \\
$D^1_{11}$ & $-\frac{1}{3}$ & $-\frac{1}{3}$ &
$-\frac{2}{3}$ & $a^2$ & $D^1_{01}$ & 0 &
$-\frac{1}{3}$ & $-\frac{1}{3}$ & $ac$ &
$D^1_{-11}$ & $\frac{1}{3}$ & $-\frac{1}{3}$ & 0 &
$c^2$ \\
$D^1_{10}$ & $-\frac{1}{3}$ & 0 & $-\frac{1}{3}$ &
$ab$ & $D^1_{00}$ & 0 & 0 & 0 & $ad+bc$ &
$D^1_{-10}$ & $\frac{1}{3}$ & 0 & $\frac{1}{3}$ &
$cd$ \\
$D^1_{1-1}$ & $-\frac{1}{3}$ & $\frac{1}{3}$ & 0 &
$b^2$ & $D^1_{0-1}$ & 0 & $\frac{1}{3}$ &
$\frac{1}{3}$ & $bd$ & $D^1_{-1-1}$ & $\frac{1}{3}$
& $\frac{1}{3}$ & $\frac{2}{3}$ & $d^2$
\end{tabular}
\end{center}
\end{table}

By Table (6.3) there are two preons, $a$ and $b$, charged and neutral,
respectively, and their respective antiparticles,  $d$ and $c$.  Other
particles may be regarded as built out of the preons  $(a,b,c,d)$ since
the values of  $(t_3 , t_0 , Q)$  of all these composite particles
may be obtained by adding the  $(t_3 , t_0 , Q)$  of each of the
constituent preons. Therefore the factors  $D_{-3t_3 -3t_0}^{3t}$  may
be read in two ways:  (a) as describing creation operators for quantum
knots representing the internal state of a composite particle or  (b) as
a product of creation operators for the component preons.

The preceding remarks are illustrated in Tables (5.1), (5.2), (6.3) and
(6.4) where  $D_{-3t_3 -3t_0}^{3t} $  and  $(Q,t,t_3, t_0)$  are
summarized for preons, bosonic preons, elementary fermions, and weak bosons.

To show that the preon interpretation holds in all representations let
us rewrite (2.1) by introducing  $(n_a , n_b, n_c , n_d)$  the exponents
of  $(a,b,c,d)$:

\be
\begin{array}{l}
n_a=s \\
n_b= n_+ -s 
\end{array}
\ee

\be
\begin{array}{l}
n_c=t  \\
n_d= n_- -t 
\end{array}
\ee

\noindent
Then

\be
\begin{array}{l}
n_+ =n_a + n_b \\
n_- =n_c + n_d 
\end{array} 
\ee

\be
\begin{array}{l}
n_+^\prime =n_a+n_c  \\
n_- ^\prime =n_b +n_d 
\end{array} 
\ee

\noindent
and
\be
\begin{array}{l}
n_a + n_b + n_c + n_d = n_+ + n_- = 2j (=N=6t)
\end{array}
\ee
\be
\begin{array}{l}
n_a + n_b - n_c - n_d = n_+- n_- = 2m (= w=-6t_3) 
\end{array}
\ee
\be
\begin{array}{l}
n_a + n_c - n_b - n_d = n_+^\prime - n_- ^\prime = 2m' (=r+1 = -6t_0 )
\end{array} 
\ee

\noindent
In the preon interpretation of $D_{mm'}^j$  the  $(a,b,c,d)$  are
regarded as creation operators for $(a,b,c,d)$ particles and the  $(n_a,
n_b, n_c, n_d )$  are the numbers of  $(a,b,c,d)$  preons in each term.  
These will vary from term to term but the left sides of Eqs. 
(6.5)-(6.7)
remain the same in all terms contributing to  $D_{mm'}^j$  and
they also have simple meanings.

Eqs. (6.5)-(6.7) may be rewritten as

\be
\begin{array}{l}
t= \frac16 (n_a + n_b + n_c + n_d ) \\
\end{array}
\ee

\be
\begin{array}{l}
t_3 = -\frac16 (n_a + n_b - n_c - n_d ) \\
\end{array}
\ee

\be
\begin{array}{l}
t_0 = -\frac16 (n_a - n_b + n_c - n_d ) 
\end{array}
\ee

\noindent
By (4.9), (6.9) and (6.10)
\[
Q= -\frac e3 (n_a - n_d )
\]

\noindent
The representations already considered, (Tables (5.1), (5.2), and (6.3))
illustrate special cases of these general relations that express  
$(t,t_3 , t_0 )$  of a composite particle in terms of the charges 
and hypercharges of the
preonic constituents.  In general when $D^j_{mm}$ is not a monomial,
the composite particle represented by $D^{3t}_{-3t_0-3t_0}$ is a
superposition of distinct structures, all having the same
$(t,t_3,t_0)$ but with varying numbers $(n_a,n_b,n_c,n_d)$ of preons.

\section{The Complementary Models.${}^9$}

The Eqs. (6.5), (6.6), (6.7), may also be read as knot relations

\be
\begin{array}{l}
n_a + n_b + n_c + n_d = N
\end{array}
\ee
\be
\begin{array}{l}
n_a + n_b - n_c - n_d = w 
\end{array}
\ee
\be
\begin{array}{l}
n_a + n_c - n_b - n_d = r+1 
\end{array}
\ee

Eq. (7.1) states that the total number of preons equals the number
of crossings $(N)$.  Since we shall assume that the preons are fermions,
the knot is a fermion or boson depending on whether the number of
crossings is odd or even.

The meaning of (7.2) and (7.3) becomes clearer if we note that $a$ and
$d$  are antiparticles since they have opposite charge and hypercharge,
while  $b$  and  $c$  are neutral antiparticles with opposite values of
the hypercharge.  We may therefore introduce the "preon numbers":

\be
\begin{array}{l}
\nu _a = n_a - n_d 
\end{array}
\ee
\be
\begin{array}{l}
\nu _b = n_b - n_c 
\end{array}
\ee

\noindent
Then  (7.2)  and (7.3) may be rewritten as

\be
\begin{array}{l}
\nu _a + \nu _b = w  ~~~ (= -6t_3)
\end{array}
\ee

\be
\begin{array}{l}
\nu _a - \nu _b = r+1 ~~~(= -6t_0) 
\end{array}
\ee

\noindent
By (7.6) and (7.7) the conservation of writhe and rotation, 
or the conservation of charge and hypercharge, is equivalent
to the conservation of the preon numbers.

These considerations have led us to the position that the symmetry of a
solitonic elementary particle, that is described by representations of
the  $SL_q(2)$  algebra, may be expressed in any of the following ways:

\be
D_{mm'}^j =  D_{-3t_3 -3t_0}^{3t} = D_{\frac w2 \frac{r+1}2}^{N/2} =
\hat D_{\nu _a \nu _b}^{N^\prime} 
\ee

\noindent
where  $N^\prime$  is the total number of preons.

We  interpret the different forms of  $D_{mm'}^j$  as showing that
different aspects of the solitonic particle all display the same
$SL_q(2)$ symmetry.

In terms of  $(N,w,r)$

\be
D_{\frac w2 \frac{r+1}2} ^{N/2} (q|abcd) =
\left[ \frac{\langle n_+^\prime \rangle_{q_1}! \langle n_-^\prime \rangle_{q_1}
!}{\langle n_+\rangle_{q_1}! \langle n_- \rangle_{q_1}! } \right] ^{1/2} \
\sum _{\scriptstyle 0\leq s\leq n_+\atop\scriptstyle 0\leq t \leq n_- }
\left< {\scriptstyle n_+ \atop \scriptstyle s}
\right> _{q_1} \left< {\scriptstyle n_-\atop \scriptstyle t} \right> _{q_1} \
\delta (s+t, n_+^\prime )a^s b^{n_{+} -s} c^t d^{n_- -t} 
\ee
\noindent
where

\be
\begin{array}{l}
n_\pm = \frac12 [N\pm w]
\end{array}
\ee
\be
\begin{array}{l}
n_\pm ^ \prime = \frac12 [N\pm (r+1)]
\end{array}
\ee
\noindent
The complementary description expressed in terms of the population
numbers  $(n_a, n_b, n_c, n_d)$  is
\be
D_{mm'}^j = \hat D_{\nu _a \nu_b }^{N^\prime} 
\ee
\noindent
where

\be
\hat D _{\nu _a \nu _b}^{N^\prime} =
\left[ \frac{\langle n_a + n_c\rangle_{q_1}! \langle n_b + n_d \rangle_{q_1}!}
{\langle n_a + n_b \rangle_{q_1}! \langle n_c + n_d \rangle_{q_1}!} \right] ^{1/2} \
\sum_{\scriptstyle N^\prime \geq n_a,n_b\geq 0 \atop \scriptstyle N^\prime \geq n_c,n_d\geq 0}
\left< {\scriptstyle n_a+n_b \atop \scriptstyle n_a }\right> _{q_1}
\left< {\scriptstyle n_c + n_d \atop \scriptstyle n_c} \right> _{q_1} \
a^{n_a} b^{n_b} c^{n_c} d^{n_d}
\ee

These complementary representations (7.9) and (7.13) are related by
\be
\hat D _{\nu _a \nu _b}^{N^\prime} = \sum _{Nwr} \delta (N^\prime , N) \delta (\nu
_a + \nu _b , w) \delta (\nu _a - \nu _b ,r+1)\
D_{\frac w2 \frac{r+1}2 } ^{N/2}
\ee
\noindent
where  $N'$  is the number of preons and  $N$  is the number of crossings.

For the fundamental and adjoint representations we have  $j=1/2$  and  
$j=1$ respectively and therefore  $N=1$  or  $N=2$, where  $N$  is the
number of crossings.  These do not describe knots,
but they do describe twisted loops.  
We may still compute  $w$  and  $r$  in the same way as for
knots.  Although these twisted loops would not have the topological
stability of knots, they could be prevented from unrolling by a
dynamical stability of  $w$  and  $r$  or equivalently by the
conservation of the preon numbers.

Viewed as a knot, a fermion becomes a boson when the number of
crossings is changed by adding or subtracting a curl.  This picture is
consistent with the complementary view of a curl as an opened preon loop.

\section{Gluon Charge.${}^9$}

The previous considerations are based on electroweak physics.  To
describe the strong interactions it is necessary according to standard
theory to introduce  $SU(3)$ charge.  We shall therefore assume that
each of the four preon operators appears in triplicate $(a_i, b_i, c_i,
d_i )$ where  $i= R,Y,G$, without changing the algebra $(A)$.  These
colored preon operators provide a basis for the fundamental
representation of  $SU(3)$  just as the colored quark operators do in
standard theory.  To adapt the electroweak operators to the requirements
of gluon fields we make the following replacements:
\begin{eqnarray}
{\rm leptons}:  & &a^3 \to \epsilon^{ijk}a_ia_ja_k \\
{\rm neutrinos}:  & & c^3 \to \epsilon^{ijk}c_ic_jc_k \\
{\rm down~quarks}:  & & a_ig^{k\ell}b_kb_\ell \\
{\rm up~quarks}:  & & c_ig^{k\ell}d_kd_\ell
\end{eqnarray}
\noindent
where  $g^{jk}$ is the group metric of  
$SU(3)$.  Here  $(i,j,k)=(R,Y,G)$  and  $(a_i b_ic_id_i)$ are
creation operators for colored preons.  Then the leptons and neutrinos
are color singlets while the quark states correspond to the fundamental
representation of  $SU(3)$, as required by standard theory.  

\section{The Elementary Fermions as Preonic Trefoils${}^9$}

Since the number of crossings equals the number of preons, one may
speculate that there is one preon at each crossing if both preons and
crossings are considered pointlike.  If the pointlike crossings are
labelled  $(\ax _1 \ax _2 \ax _3 )$, then by  (8.1)-(8.4) the wave
functions of the trefoils representing leptons  $(\ell)$, neutrinos  
$(\nu)$, down quarks  $(d)$, up quarks  $(u)$  are as follows:

\be
\begin{array}{l}
\Psi _\ell (\ax _1 \ax _2 \ax_3) = \epsilon ^{ijk} \psi _i (a|\ax _1 )
\psi _j (a|\ax _2 ) \psi _k (a|\ax _3 ) 
\end{array}
\ee
\be
\begin{array}{l}
\Psi _\nu (\ax_1 \ax_2 \ax _3) = \epsilon ^{ijk} \psi _i (c|\ax _1 )
\psi _j (c|\ax _2 ) \psi _k (c|\ax _3 ) 
\end{array}
\ee
\be
\begin{array}{l}
\Psi _d (\ax _1 \ax _2 \ax _3 ) = \psi _i (a|\ax _1 ) g^{jk}\psi_j(b|
\vec x_2)\psi_k(b|\vec x_3)
\end{array}
\ee
\be
\begin{array}{l}
\Psi _u (\ax _1 \ax _2 \ax _3 ) = \psi _i (c|\ax _1 )  g^{jk}\psi_j(d|\vec x_2) \psi_k(d|\vec x_3) 
\end{array}
\ee

\noindent
where  $i= (R,Y,G)$  and  $\psi _i (a|\ax ) \dots \psi _i (d|\ax )$  are
colored  $\delta$-like functions localizing the preons at the crossings.

Then the wave function of a lepton describes a {\it singlet} trefoil
particle containing three preons of charge  $(-e/3)$  and hypercharge  
$(-e/6)$.  The corresponding characterization of a neutrino describes a
{\it singlet} trefoil containing three neutral preons of hypercharge  
$(-e/6).$

The wave function of a down quark describes a colored trefoil particle
containing one $a$ preon with charge  $(-e/3)$ and hypercharge  $(-e/6)$  
and two neutral  $b$ preons with hypercharge  $(e/6)$.  The
corresponding characterization of an up-quark describes a colored
trefoil containing two charged $d$ preons with charges  $(e/3)$  and
hypercharge  $(e/6)$ , and one neutral $c$ preon with hypercharge  $(-e/6)$.

This hypothetical structure is held together by the trefoil of fields
connecting the preons.  A search for this kind of substructure
depends critically on the mass of the conjectured preons and the
strength with which they are bound. 

\section{Preons as Physical Particles.${}^4$}

We have so far viewed the preons mainly as a simple way to describe the
algebraic structure of the knot polynomials.  If these preons are in
fact physical particles, the following decay modes  of the quarks are
possible.

\[
\mbox{Down quarks:}~~~~ {\cal D} _{\frac 32 - \frac 12}^{3/2} \rightarrow
{\cal D} _{ \frac 12 \frac12 }^ {1/2} + {\cal D} _{1-1} ^1, \;\;  (ab ^2
\rightarrow a+b^2 )
\]
\noindent
or
\[
\mbox{Up quarks:}~~~~~ {\cal D} _{-\frac 32 - \frac 12}^{3/2} \rightarrow
{\cal D} _{- \frac 12 \frac12 }^ {1/2} + {\cal D} _{-1-1} ^1, \;\; (cd ^2
\rightarrow c+d^2 )
\]
\noindent
and the preons could play an intermediary role as virtual particles in
quark processes.

The simple knot model predicts an unlimited number of excited
states${}^{8}$ but it appears that there are only three generations,
e.g. $(d,s,b)$.  According to the preon scenario, however, it may be
possible to avoid this problem by showing that the quarks will
dissociate into preons if given a critical ``dissociation energy'' less
than that needed to reach the level of the fourth predicted flavor.  In
that case one would also expect the formation of a preon-quark plasma at
sufficiently high temperatures.

It may be possible to study the thermodynamics of the plasma composed of
quarks and these hypothetical particles.

Since the  $a$  and  $d$  particles are charged  $(\pm e/3)$  one
should expect their electro-production according to
\[
e^+ + e^- \rightarrow a+ \bar a \dots
\]
\noindent
at sufficiently high energies of a colliding  $(e^+ , e^- )$  pair.

If the preons are assumed to be pointlike they must also
be very heavy.  If the trefoil model is considered seriously for the
leptons and neutrinos, then the binding energy must nearly compensate
the mass of the very heavy constituent preons with a correspondingly
higher melting temperature for the leptons and neutrinos.



\section{Summary and Comments}

In this paper the quantum knot has been characterized kinematically
by  $D_{mm^\prime}^j (abcd)$, an element of an irreducible representation of the
knot algebra  $SL_q(2)$  with
\be
(j,m,m^\prime)= \frac12 \ (N,w,\pm r+1)
\ee

\noindent
where the spectrum of  $(j,m,m^\prime)$  is limited by  $SL_q(2)$  and the spectrum
of  $(N,w,r)$  is restricted by the topology of a classical knot.  The pair  $(w,r)$ 
are topological constants of the classical motion and the pair  $(m,m^\prime)$  are
quantum constants of the motion by virtue of the  $U_a(1) \times U_b (1)$  invariance
of the  $SL_q(2)$  algebra.  If an elementary particle is identified
a a quantum knot, it acquires in addition to the familiar angular
momentum and isotopic spin, new degrees of freedom associated with the
knot algebra as represented by $D^j_{mm^\prime}$.

When the 12 elementary fermions are described as 3 states of excitation of 4 quantum
trefoils, each quantum trefoil corresponds to one family of 3 fermions.  The
correspondence is unique and is expressed by the empirical relation
\be
(j,m,m^\prime)=3(t,-t_{3},-t_0)
\ee
\noindent
where  $(j,m,m^\prime)$  describes one of the four quantum trefoils and  $(t=1/2, t_3 ,t_0)$ 
describes one of the four fermion families (leptons, neutrinos, down quarks, up quarks).  Eq.
(11.2)  records a one-to-one correspondence between the four quantum trefoils and the
four fermion families.

In order to conserve $t_3$ and $t_0$ in all interactions the
relation (11.2) is next extended to hold for all particles.  Then for elementary
fermions  $j=3t=3/2$, and for the triplet of weak bosons 
$(W^+ W^- W^0)$, one has $j=3t=3$,  while for
the fourth weak boson  $j=3t=0$.  The particles belonging to the fundamental  $(j=1/2)$
and adjoint  $(j=1)$  representations of  $SL_q(2)$  are new particles that may be called preons,
with values of  $(t,t_3, t_0)$  given by  (11.2).

The set  $(t,t_3,t_0)$  are defined as indices of  $SL_q(2)$  but  
$Q=(t_3 + t_0)e$  and  $t_0$  have
their usual physical meaning as charge and hypercharge.  One then finds that the particles
with higher values of  $(t,t_3,t_0)$  may be regarded as built up of the four preons (one
charged, one neutral, and their antiparticles) belonging to the fundamental representation
of  $SL_q(2)$.  These composite particles composed of preons are also characterized by (11.1),
where  $(N,w,\pm r+1)$  may be interpreted to describe a quantized knotted field binding the preons
together.  Both the field and particle aspects of the composite particle express the  $SL_q(2)$
symmetry.

In this way the intuitive trefoil picture, when implemented empirically as the  $j=3/2$
representation of the knot algebra, leads naturally to the fundamental  $(j=1/2)$ representation
of  $SL_q(2)$  and the preonic constructions.  This development resembles the transition from the
``8-fold way", the adjoint representation of  $SU(3)$, to the fundamental representation of 
$SU(3)$ and the quark constructions.

An unsatisfactory feature of the model, however, is the meaning of
$q$, which is obscure.  Like
Planck's constant, which normalizes the non-Abelian Heisenberg algebra, the parameter  $q$  also
normalizes a non-Abelian algebra, 
but an algebra dependent on  $\epsilon_q$  
\break instead of  $i$  where  
$\epsilon_q$
is a different square root of  $-1$.  Unlike  $h$, which has the dimensions \break of an action, $q$
is dimensionless.


The introduction of substructure, determined by the $SL_q(2)$
algebra, for the quantum fields in terms of preons
resembles the introduction of substructure for the quantum fields
in terms of field quanta determined by the Heisenberg algebra holding
for conjugate field operators.  This analogy suggests a comparison
of the $SL_q(2)$ algebra, determined by $q$, with the Heisenberg
algebra, determined by $h$ and may be based on the following quadratic
form$^{10}$ invariant under $SL_q(2)$ transformations:
\be
K=A^t \varepsilon_{q} A 
\ee

\noindent
where
\be
\varepsilon _q = 
\left( 
\begin{array}{cc}
0 & q^{-1/2} \\
-q^{1/2} &0
\end{array}
\right)  \;\;\;\;\;\;\;\;\;\; \varepsilon_q^{2} = -1
\ee
\noindent
$K$  is invariant under  $SL_q(2)$  transformations of  $A$:

\be
A^\prime= T\ A \qquad \;\;\;\;\;\; T\varepsilon SL_q(2) 
\ee

\noindent
Choosing

\be
A= \left(
\begin{array}{c}
D_{x} \\
x
\end{array}
\right)
\ee

\noindent
and normalizing

\be
K=q^{-1/2}  
\ee

\noindent
one has by (11.3) the following $SL_q(2)$ invariant relation  

\be
D_x x-q x  D_x =1 
\ee
\noindent
Equation 
(11.8)  is satisfied if  $D_x $  is chosen as the  $q$-difference operator, namely
\be
D_x \psi(x) = \frac{\psi (qx) - \psi (x)}{ qx - x}  
\ee
If we introduce
\be
P_x = \frac{\hbar}i \ D_x 
\ee
then (11.8)  becomes
\be
(P_x x-q x P_x) \psi (x) = \frac{\hbar}i \ \psi (x) 
\ee
\noindent
If  $q\rightarrow 1$, then (11.11) becomes the Heisenberg commutator applied to a quantum
state.  If  $q$  is near unity (as it must be insofar as the standard theory $(q=1)$  is
approximately correct)  then  
by (11.9) $D_x$  resembles the differentiation operator on a lattice
space and  $q$  may play the role of a dimensionless regulator.

In view of the physical evidence suggestive of substructure, which has been described
here, as well as the natural appearance of the non-standard 
$q$-derivative, it may
be possible to utilize  $SL_q(2)$  to describe a finer level of structure than is
currently considered.

We have ignored the gravitational field in this paper since it is not immediately
relevant.  As we have, however, discussed the knot symmetrices of the fundamental
particles, we have thereby also discussed the knot symmetries of these sources of the
gravitational field.  Since one expects that the symmetries of its source would in
some measure be inherited by the gravitational field itself, it is interesting that
knot states have emerged in a natural way from and are therefore compatible with
attempts to quantize general relativity.${}^{11}$

\vskip.5cm

\section{Appendix$^{12}$}

Here we show the invariance of the modified action under the complete
gauge group
\be
{\cal{S}} = S\times s
\ee
where
\begin{eqnarray*}
S &=& \mbox{local}~ SU(2)\times U(1) \\
s &=& \mbox{global}~ U_a(1) \times U_b(1)
\end{eqnarray*}
The vector potential of the non-Abelian part of the standard model is
\be
\begin{array}{rcl}
{\cal{W}}_\mu &=& i~gW_\mu^kt_k ~~~ k = (+,-,3) \\
t_\pm &=& \frac{1}{2} (\sigma_1 + i\sigma_2) \\
t_3 &=& \sigma_3
\end{array}
\ee
The corresponding vector potential of the modified standard model may
be chosen as
\be
{\cal{W}}_\mu = i~gW_\mu^k \tau_k
\ee
where, in the notation of Table (4.2),
\be
\begin{array}{rcl}
\tau_k &=& c_kt_kD_k ~~~ k = (+,-,3) \\
D_+ &=& D^3_{-30}/{\cal{A}}_+ = \bar b^3\bar a^3 \\
D_- &=& D_{-30}^3/{\cal{A}}_- = a^3b^3 \\
D_3 &=& D^3_{00} = f(\bar bb)
\end{array}
\ee
and where the $c_k$ are free numerical constants.  Here ${\cal{A}}_+$
and ${\cal{A}}_-$ are the numerical factors appearing in (2.1).  The
covariant derivative and field strength are constructed in the
familiar way:
\begin{eqnarray}
\nabla_\mu &=& \partial_\mu + {\cal{W}}_\mu \\
{\cal{W}}_{\mu\lambda} &=& [\nabla_\mu,\nabla_\lambda]
\end{eqnarray}
Then one may show that the modified field strength is$^{12}$
\be
{\cal{W}}_{\mu\lambda} = W^s_{\mu\lambda}\tau_s + \hat W^s_{\mu\lambda}
D_s
\ee
where
\be
W^s_{\mu\lambda} = ig(\partial_\mu W^s_\lambda - \partial_\lambda
W^s_\mu) -g^2 f^s_{m\ell}(\bar bb)W^m_\mu W^\ell_\lambda
\ee
and
\be
\hat W^s_{\mu\lambda} = -\frac{1}{2} g^2\delta(\ell,\pm)
\delta(m,\mp) \hat f^s_{m\ell}(\bar bb) W_\mu^m W_\lambda^\ell
\ee

The modified structure constants $f^s_{m\ell}(\bar bb)$ and
$\hat f^s_{m\ell}(\bar bb)$ become numerical when evaluated on the
ground state $|0\rangle$ of the $q$-oscillator.

The $\hat f^s_{m\ell}$ vanish unless $(m,\ell) = (\pm,\mp)$ and
$s=3$, and therefore $\hat W^s_{\mu\lambda}$ also vanishes unless
$s=3$.

We choose the modified field action to be
\be
\langle 0|\rm{Tr}~{\cal{W}}_{\mu\lambda}{\cal{W}}^{\mu\lambda}|0\rangle
\ee
where the trace is over the $t_k$ matrices and $|0\rangle$ is the 
ground state of the $q$ oscillator.  The structure constants
$f^s_{k\ell}(\bar bb)$ and $\hat f^s_{k\ell}(\bar bb)$ become
numerical in (12.10).

After reducing the trace one finds
\be
\rm{Tr}~{\cal{W}}_{\mu\lambda}{\cal{W}}^{\mu\lambda} =
W^m_{\mu\lambda}W^{p\mu\lambda}\rm{Tr}~\tau_m\tau_p + 2
\hat W^m_{\mu\lambda}\hat W^{p\mu\lambda}D_mD_p
\ee
and
\be
{\cal{S}}(\rm {Tr}~{\cal{W}}_{\mu\lambda}{\cal{W}}^{\mu\lambda})
{\cal{S}}^{-1} = c_mc_p W^m_{\mu\lambda}W^{p\mu\lambda}~\rm{Tr}~
t_mt_p(sD_mD_ps^{-1} + 2\hat W^m_{\mu\lambda}\hat W^{p\mu\lambda}
(sD_mD_ps^{-1})
\ee
where ${\cal{S}} = S\times s$ and there is the standard invariance
under $S$.  The factors $sD_mD_ps^{-1}$ remain to be considered.
In the first term on the right Tr $t_mt_p$ vanishes unless
\be
(m,p) = (\pm,\mp) ~~~\mbox{or} ~~~ (m,p) = (3,3)
\ee
In both cases $D_mD_p$ is neutral and therefore $sD_mD_ps^{-1} =
D_mD_p$.  The second term on the right vanishes by (12.9)
 unless
\be
(m,p) = (3,3)
\ee
Again, since $D_mD_p$ is neutral, this term is also invariant.
Therefore the modified self-interaction of the non-Abelian vector field
remains invariant under $U_a\times U_b$ as well as under $S$.

To show the invariance of the modified Fermion-Boson interaction
under $U_a\times U_b$, one also needs to consider the modification
introduced by the following form factor
\be
\bar D^{j^{\prime\prime}}_{m^{\prime\prime}p^{\prime\prime}}
D^j_{mp}D^{j^\prime}_{m^\prime p^\prime}
\ee
which multiplies the term that previously described the standard
Fermion-Boson interaction.  We shall impose $U_a(1)\times U_b(1)$
invariance on this form factor and hence on the modified interaction.

By (2.5) this invariance requires in (12.15):
\be
(m,p) = (m^{\prime\prime},p^{\prime\prime})-(m^\prime,p^\prime)
\ee
But since the initial $(m^\prime,p^\prime)$ and final states 
$(m^{\prime\prime},p^{\prime\prime})$ represent fermions, 
the boson $(m,p)$ is by (4.15)
\be
(m,p) = -3[(t_3,t_0)^{\prime\prime}-(t_3,t_0)^\prime]
\ee
and since $(t_3,t_0)$ are conserved in the standard model one has
\be
(m,p) = -3(t_3,t_0)
\ee
for the intermediate boson as well as for the fermions.
Eqs. $(4.17m)$ and $(4.17m^\prime)$ are therefore necessary 
conditions for $U_a(1)\times U_b(1)$ invariance in this model.
Also in (12.15)
\[
j^\prime + j^{\prime\prime} \geq j \geq |j^\prime-j^{\prime\prime}|
\]
and since $j\geq |m|$,
\[
D^j_{\pm 30} = D^3_{\pm 30}
\]
Therefore the invariance of (12.15) is not only required for the
existence of the charges $Q_a$ and $Q_b$ but also implies
\[
(j,m,m^\prime) = 3(t,-t_3,-t_0)
\]
for the intermediate charged boson.

\vskip.5cm

\noindent
{\bf Acknowledgement:}  I thank J. Smit and A. Cadavid for helpful discussions.

\vskip0.5cm
\noindent
{\bf  References}

\begin{enumerate}
\item
Thomson, W.H., {\it Trans. R. Soc. Edinb}. {\bf 25}, 217-220 (1969).
\item
Faddeev, L. and Niemi, Antti{\it J. Nature} {\bf 387}, May 2 (1997).
\item
Pati, J. and Salam, A., {\it Phys. Rev. D} {\bf 10}, 275-289 (1974).
\item
Finkelstein, R.J., {\it Int. J. Mod. Phys.} A{\bf 24}, 2307 (2009).
\item
Reshetikhin, N.YU, Takhtadzhyan, L.A. and Faddeev, L.D., {\it Leningrad Math. J.} {\bf 1} (1990).
\item
Finkelstein, R.J., {\it Int. J. Mod. Phys.} A{\bf 22}, 4467 (2007).
\item
Kauffmann, L.H., in {\it Quantum Groups}, eds. T. Curtright et al World Scientific (1991).
\item 
Cadavid, A.C. and Finkelstein, R.J., {\it Int. J. Mod. Phys.} A{\bf 21} 4269 (2006).
\item
Finkelstein, R.J.,  arXiv:0901.1687.
\item
Finkelstein, R.J. and Marcus E., {\it J. Math Phys.}, A{\bf 21}, 4269 (2006).
\item Horowitz, G.T., in {\it Strings and Symmetries}, World Scientific
(1991).
\item Finkelstein, R.J., arXiv:hep-th/0701124, v1.
\item Finkelstein, R.J., arXiv:hep-th/1011.2545.
\end{enumerate}

\end{document}